\documentclass[review,number,sort&compress]{elsarticle}

\usepackage{amsmath}
\usepackage{graphicx}
\usepackage{multirow}
\usepackage{multicol}
\usepackage[english]{babel}
\usepackage{version}
\usepackage{fullpage}
\usepackage{hyperref}
\usepackage{subfigure}
\usepackage{color}
\begin{document}
   
\title{Study and optimization of the spatial resolution for detectors with binary readout}
\author{R. Yonamine}
\ead{ryo.yonamine@ulb.ac.be}
\author{T. Maerschalk}
\author{G. De Lentdecker}
\address{Universit\'{e} Libre de Bruxelles, 1050 Bruxelles, Belgium}

\begin{abstract}
Using simulations and analytical approaches, we have studied single hit resolutions obtained with a binary readout,
which is often proposed for high granularity detectors to reduce the generated data volume.
Our simulations considering several parameters 
(e.g. strip pitch)
show that the detector geometry and an electronics parameter of the binary readout chips could be optimized for binary readout 
to offer an equivalent spatial resolution to the one with an analogue readout. 
To understand the behavior as a function of simulation parameters, we developed analytical models
that reproduce simulation results with a few parameters.
The models can be used to optimise detector designs and operation conditions
with regard to the spatial resolution. 
\end{abstract}


\maketitle

\section{Introduction}

In large experiments, comprising hundreds or thousand of detection elements, 
it is sometimes more advantageous to use a binary readout electronics than an analogue one. Indeed it is not always feasible to integrate an analog-to-digital converter in each channel of the front-end ASIC; the constraints are typically associated with the total area of the integrated circuit and with the power consumption. With a binary readout the cost of an increased number of readout channels would then be balanced by a simpler readout circuitry. In addition the data volume is much smaller with a binary readout.\\
For many applications one can use the binary readout architecture. In this architecture each channel of the front-end electronics is equipped with an amplitude discriminator which generates 1-bit information in response to each signal above a given threshold. The information delivered by a strip detector is suppressed to the minimum already in the front-end circuit. Binary information can be easily stored in the integrated circuit separately for each channel, which allows one to cope with high rate of particles.

Another important concern for tracking detectors is
the spatial resolution;
however, it is not trivial if signal charges are shared
with more than one readout strip.
In this paper, we estimate by Monte Carlo simulations and analytically 
the spatial resolution with a binary readout for three types of detectors: silicon sensor, Micromegas, and GEM-based detectors. 

This paper is organized as follows.
Sec.\,\ref{sec:detector_model} introduces a detector model and the technologies we refer to in this paper.
Sec.\,\ref{sec:simulation} describes how our simulations work and presents a simulation result,
followed by discussions with our analytical models in Sec.\,\ref{sec:analytical}.
A conclusion is given in Sec.\,\ref{sec:conclusion}.

\section{Detector Model}
\label{sec:detector_model}
\begin{figure}[!ht]
     \begin{center}
      \includegraphics[width=0.6\textwidth]{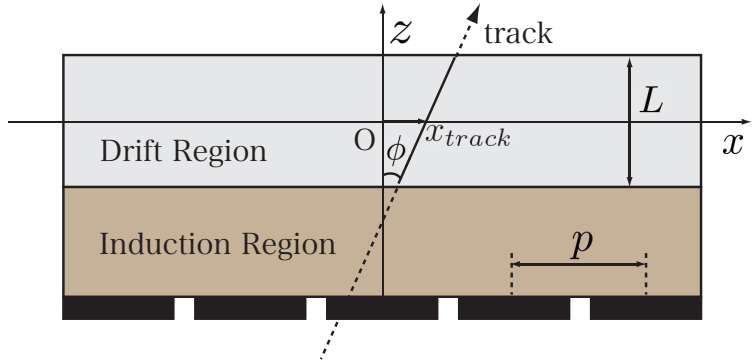}
      \caption{A cross section of the detector model. In this report, only $\phi=0$ condition is considered.}
      \label{geom}
      \end{center}
    \end{figure}

Let us begin by describing a simple detector model with some parameters
with which different types of detector technologies are distinguished from each other.
The detector model consists of the drift region and the induction region,
which are separated by an amplification step
(Fig.\,\ref{geom}).
The drift region is the sensitive part of the detector and the 
induction
region is the volume where the signal is induced to the electronics. The drift region and the 
induction
region have a parametrized size. 
The other end of the induction region is equipped with
the electrode strips with the pitch $p$. The strips are to be connected to the readout electronics.

Those detectors are used to reconstruct the position of a `charged particle track'. 
The position of a track is defined as the midpoint of the track segment in the drift region and is referred to $x_{track}$ in this study.
Note that $x_{track}$ can be always defined from the center of the nearest strip to the track position and thus $-p/2\leq x_{track} \leq +p/2$. 

The parameters used to compute the spatial resolution in this study are the number of primary electrons and the gap of drift region $L$, the gap of the induction region, 
the track incident angle $\phi$ from the vertical axis ($z$) to the strip plane,
the transverse diffusion coefficient 
$C_d:=\sigma_{\rm d}/\sqrt{z}$ with $\sigma_{\rm d}$ being the diffusion width of an electron cloud which has drifted over a length $z$,
namely $C_d^{\rm dr}$ for the drift region and $C_d^{\rm in}$ for the induction region,
and an electronics threshold.

\subsection{Micromegas}

\begin{figure}[!ht]
\begin{center}
\includegraphics[scale=0.55]{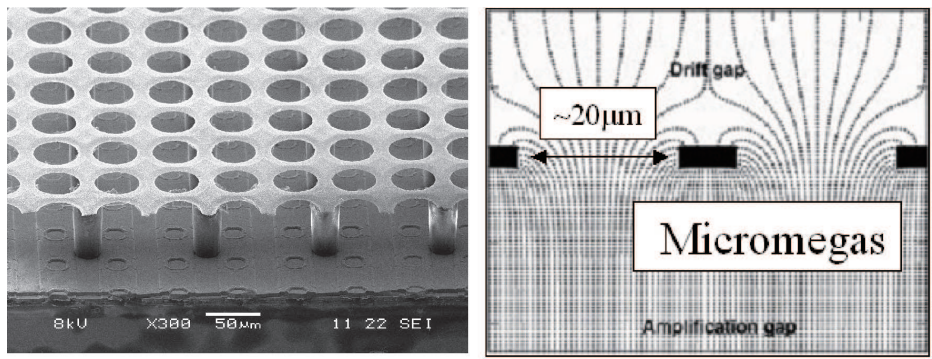} 
\caption{SEM image of a Micromegas structure (Left)\cite{Melai:2010sg}.The electric field lines around the Micromegas grid (Right).}
\label{mesh}
\end{center}
\end{figure}

The Micromegas (MICRO MEsh GAseous Structure) concept was introduced in 1996 by I. Giomataris et al.\cite{Giomataris:1995fq}.
The detector is filled with gas mixtures and charged particles create
seed electrons, which will contribute to inducing signals, along their trajectories by ionizing the gas molecules.
The created electrons in an electric field drift toward the anode plane,
above which a cathode grid is placed (Fig.\,\ref{mesh}). 
This grid is maintained at $100$ microns of the anode plane. 
A voltage of typically $-400$\,V is applied to the grid and the anode plane is grounded via the readout electronics to create an electric field of about $40$\,kV/cm. This space, between the grid and the anode plane, define the volume of amplification
which coincides with the induction region, in this case. \\

\subsection{GEM}

\begin{figure}[!ht]
\begin{center}
\includegraphics[scale=0.8]{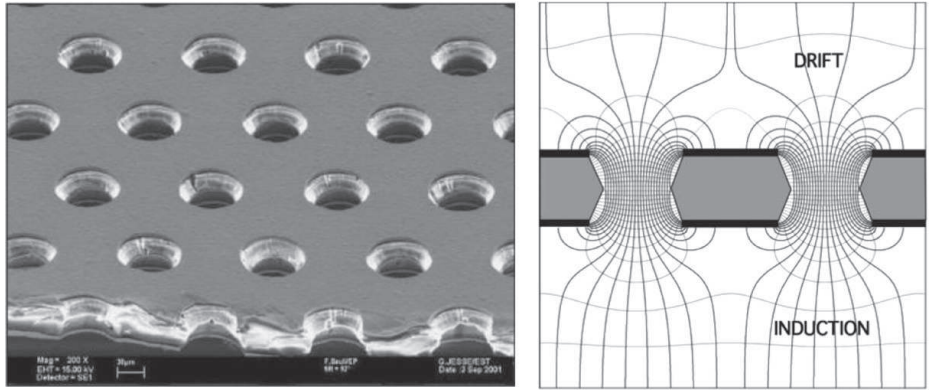} 
\caption{SEM image of a GEM surface (Left) and the electric field lines (Right)\cite{Sauli20162}.}
\label{GEM}
\end{center}
\end{figure}

The GEM (Gas Electron Multiplier) concept was proposed in 1997 by F. Sauli~\cite{Sauli:1997qp}. 
The detector has the same sensitive volume as the Micromegas but has a different amplifying structure.
The GEM consists of a 50~$\mu$m Kapton foil, copper cladded on both sides, chemically pierced with a high density of micro-holes (typically $50$ to $100$ holes per cm$ ^ 2 $).
The holes 
have a
bi-conical shape and have a diameter of about $ 50$\,$\mu$m. Fig.\,\ref{GEM} shows a SEM image of a GEM surface with its dimensions (Left) and the electric field lines in GEM holes (Right).
The detector is typically built with multiple GEM layers (generally two or three) to achieve stable amplification
or to reduce ion back-flow rate, which ions degrade detector performance. 

\subsection{Semiconductor detectors}
\begin{figure}[!ht]
     \begin{center}
      \includegraphics[width=0.6\textwidth]{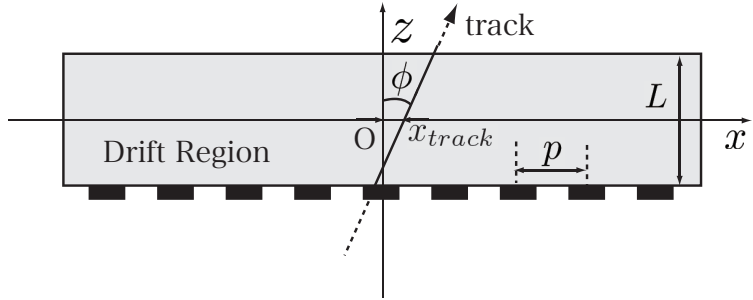}
      \caption{Schematic drawing of semiconductor detectors.}
      \label{geomSi}
      \end{center}
    \end{figure}

The most important difference compared to Micromegas and GEM-based detectors is 
that a silicon detector uses a semiconductor material for its sensitive volumes instead of gas mixtures.
Since the semiconductor has a factor 10 smaller ionization potential than gas,
modern low-noise electronics can read out signals without amplification contrary to the gas-based detectors.\\
There is neither amplification nor induction region. The induction region coincides with the sensitive region also called the drift region (see our detector model above).
Fig.\,\ref{geomSi} shows a schematic drawing of a semiconductor detector.

\section{Simulation}
\label{sec:simulation}

A simulation of the detector has been developed following the model described before. 
We fix a simple geometry (the gap sizes, the pitch, and diffusion coefficients, etc.), then we create a track. 
Electrons are created along the track in the drift region. 
Those electrons drift and diffuse until the amplification region. 
Each electron arriving at the amplification region is multiplied by the gain $g$. 
After amplification, all electrons drift and diffuse until the strip plane.\\
We will now describe in details the implementation of all those steps in the simulation.

\subsection{Charged particle track}

\begin{figure}[!ht]
     \begin{center}
      \includegraphics[width=0.7\textwidth]{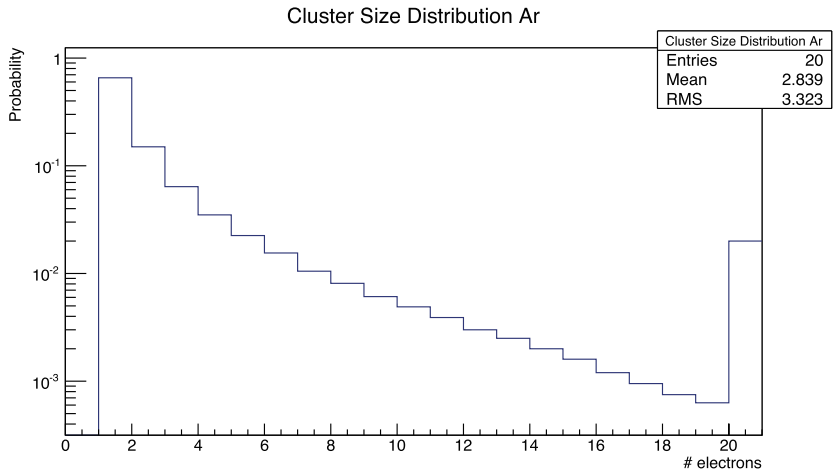}
      \caption{Argon ionization cluster size distribution is made from data comming from Ref.\,\cite{Blum:1993}.}
      \label{IClz}
      \end{center}
    \end{figure}

We will 
first discuss 
the case of incident tracks perpendicular to the strip plane ($\phi = 0^{\circ}$), then we will discuss the angular case.\\
The straight track is defined by a number of electron clusters uniformly distributed in the drift region 
 and with $x = x_{track}$. \\
It is well known that the detector medium affects the
total number of ionized electrons,
for instance, $\sim$100/1cm for MIP in Ar gas,
$\sim$10,000/100$\mu$m for MIP in a silicon sensor. To take this fact into account in the simulation, the parameters for the gas-based and the semiconductor detectors are not the same.\\
For the gas-based detectors the number of clusters is taken randomly on a Poisson distribution with a mean value equal to 12,
which is obtained from {\it Magboltz}~\cite{magboltz} for a 3~mm argon based gas mixture.
Each cluster has a certain number of electrons. 
The number of electrons is taken randomly on the argon ionization cluster size distribution made from data comming from Ref.\,\cite{Blum:1993},
which is shown in Fig.\,\ref{IClz}. 
The last bin of this distribution is the probability to have twenty or more electrons in one cluster.\\
For the semiconductor case, the number of electrons per cluster is fixed to one, but the number of clusters is fixed to $20000$, which is a typical value for $300~\mu$m Si sensor.

For non perpendicular tracks,
the electron clusters are uniformly distributed in the drift region according to
$x = x_{track} + \tan\phi\,z$. 
By definition, the angular track will create more clusters. The number of clusters is computed in the same way as 
the straight track (Poisson distribution for the gas-based detectors and the fixed number for the semiconductor),
and then multiplied by $\dfrac{1}{\cos\phi}$. 

\subsection{Electron's motion and gain}

The motion of each electron is defined by the following method. 
We compute the distance $L$ between the position of the electron and the position of 
the end of the drift region along the $z$ axis.
Then we compute the diffusion with :

\[
\sigma _d = C_d . \sqrt{L}
\]
Where $C_d$ is the transverse diffusion coefficient. 
The new $x$ position of the electron is randomly chosen on a Gaussian with a mean of $x_{track}$ 
and a sigma of $\sigma_d$.\\
Then, in the gas based detector, to simulate the gain $g-1$ electrons are created at that new position ($x_{\rm ampl}$).\\
Each of those electrons is then moved to the readout plane and its $x$ position is computed with the same diffusion formula (with $L$ equal to the induction gap size).

\subsection{Induction and threshold}
To simulate the strips and the induced signal of the electrons on those strips, a histogram is created. 
The bin size is equal to the strip pitch, in this way, each bin represent a strip. 
For each electron, the bin corresponding to the final $x$ position of this electron is filled. When the histogram is filled, the electronics threshold is applied.\\
For each bin, the number of electrons is compared with the threshold.\\
In the case of the binary read-out, if the number of electron is below the threshold, the output is set to zero. 
If the number of electrons is above the threshold, the output is set to one.\\
In the case of the analog read-out, if the number of electron is below the threshold, the output is also set to zero. If the number of electron is above the threshold, 
the output is set to the number of electrons. 
This output gives the information of the charge.

\subsection{Reconstructed position}

With the analog output, we reconstruct the $x$ position with the center of gravity (CoG) method, and with the binary output, we use an average position of the hit strips.\\
To obtain the spatial resolution we use the residual distribution. 
The residual is the distribution of the difference between the reconstructed position and the real position (in our case, the simulated position of the particle track). 
The mean of the residual distribution is, by definition, the bias of the reconstruction method and the RMS of the distribution is used as the spatial resolution of the method.

\paragraph{Binary Read-out}
For a binary read-out chip, all adjacent strips with a collected charge above a given threshold are considered.
The reconstructed position of the charged particle track with the binary read-out ($x_{reco,bin}$) is the geometrical center of those strips:
\[
x_{reco,bin} = \dfrac{1}{N} \sum x_i,
\]
where $N$ is the number of hit strips and $x_i$ is the center of the $i$-th strip.

\paragraph{Center of Gravity}
The position reconstructed with the CoG method ($x_{reco,CoG}$) is computed using the following formula:
\[
x_{reco,CoG} = \dfrac{1}{q_{tot}} \sum q_i x_i,
\]
where $q_{tot}$ is the total number of electrons, $i$ is the strip number, $q_i$ is the number of electrons, and $x_i$ the center of the strip $i$.

\subsection{Spatial resolution}
As a matter of fact, the reconstructed position from measured data is somehow displaced
from the original track position because of some stochastic processes in detectors and the finite strip pitch size.
Therefore it is essential to take into account the spatial resolution
to reconstruct the track position from measured data.
The spatial resolution $\sigma_{x}$ can be defined with a probability $P(x_{reco};x_{track})$ 
that a position $x_{reco}$ is reconstructed for the track position $x_{track}$:
\begin{eqnarray}
\label{resoldef}
\sigma^2_{x}(x_{track}) 
& = & \int dx_{reco} \, P(x_{reco}; x_{track}) \,  (x_{reco} - x_{track})^2. 
\end{eqnarray}
Experimentally, the RMS of the residual ($x_{reco}-x_{track}$) is often used to
estimate $\sigma_{x}$.\\
One can note that this definition is completely generic. It means that the time resolution or the energy resolution can be defined with the same expression by interpreting the meaning of $x_{reco}$ and $x_{track}$ as time or energy instead of position.

\subsection{A simulation result}
\begin{figure}[!ht]
     \begin{center}
      \includegraphics[width=0.55\textwidth]{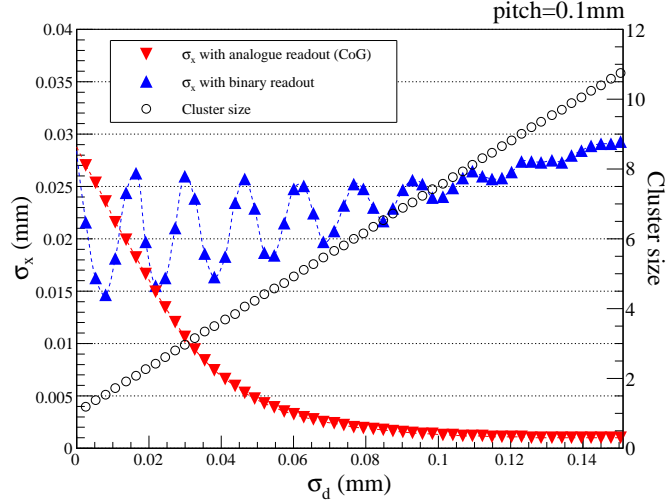}
      \caption{Spatial resolution $\sigma_x$ and the average number of hit strips (cluster size) 
               as a function of $\sigma_d$ for a typical semiconductor configuration,
               assuming the track angle $\phi=0^{\circ}$.}
      \label{sim_result}
      \end{center}
    \end{figure}

Fig.\,\ref{sim_result} shows an example of our simulation results in which spatial resolutions
and the average number of hit strips (cluster size)
are plotted as a function of $\sigma_d$ assuming typical semiconductor configurations and a track angle of $\phi=0$.
Note that each point represents a spatial resolution for a specific detector configuration that has
a specific $\sigma_d$.
The fact is that a $\sigma_d$ represents several detector cases since $\sigma_d$ depends on
the drift region gap and the bias voltage. 

The spatial resolution with the analogue readout (CoG) improves as $\sigma_d$ becomes larger,
since the CoG method does not work well when the number of hit strips is less than three,
which is described as the effects of finite size pads in Ref.\,\cite{Arogancia:2007pt}.
On the other hand, the spatial resolution with the binary readout has
two characteristics:
\begin{enumerate}
\item a wavy structure, which makes roughly twice the difference at maximum,
\item the wavy structure becomes less visible as $\sigma_d$ increases.
\end{enumerate}
The understanding of these behaviors may open the possibility to improve the spatial resolution with the binary readout.

\newpage
\section{Analytical examination}
\label{sec:analytical}
Fig.\,\ref{sim_result} shows
a detector characterized as $\sigma_d=0.016$, for example,
can be improved by a factor of $\sim$2 at maximum
if $\sigma_{d}$ becomes smaller by changing detector parameters such as an electric field.
This means it is very useful if we can see potential performances for detectors 
as can be seen in Fig.\,\ref{sim_result} to optimize the detector performance.
In this section we focus on more analytical aspects in order to 
build a method that reveals potential performances for existing detectors.
\subsection{Introduction}      
\paragraph{Accuracy term and Precision term}
The spatial resolution is defined at Equation~(\ref{resoldef}).
However, since the systematic error due to the finite strip pitch has to be taken into account, 
the definition should be written as:
\begin{eqnarray}
\sigma^2_{x} 
&:= & \frac{1}{p}\int_{-p/2}^{+p/2} dx_{track}\int dx_{reco} \, P(x_{reco}; x_{track}) \,  (x_{reco} - x_{track})^2 \cr
&=& \frac{1}{p}\int_{-p/2}^{+p/2} dx_{track} \,\left<(x_{reco} - x_{track})^2\right>,
\label{sigx2_def_tmp}
\end{eqnarray}
where we introduced the notation $\left<\,\right>$ representing 
$\int dx_{reco}P(x_{reco};x_{track})$.
We rewrite Equation~(\ref{sigx2_def_tmp}) to split the formula into two terms as:
\begin{eqnarray}
\sigma^2_{x} 
&=& \frac{1}{p}\int_{-p/2}^{+p/2} dx_{track} \,
\left[
\left<(x_{reco} - \left<x_{reco}\right>)^2\right> + (\left<x_{reco}\right> - x_{track})^2
\right]\cr
\left(\frac{\sigma_{x}}{p}\right)^2&=& \int_{-1/2}^{+1/2} d\left(\frac{x_{track}}{p}\right) \,
\left[
\left<\left(\frac{x_{reco}}{p} - \left<\frac{x_{reco}}{p}\right>\right)^2\right> + \left(\left<\frac{x_{reco}}{p}\right> - \frac{x_{track}}{p}\right)^2
\right].
\label{sigx2_def_splited}
\end{eqnarray}
In the last line, we divided both side of the equation by $p^2$ to be described by dimensionless parameters: $\sigma_x/p$, $x_{track}/p$, and $x_{reco}/p$. 
The first term is the variance of $x_{reco}/p$
and the second term is the deviation of $\left<x_{reco}/p\right>$ from the true position $x_{track}/p$,
and thus these terms will be referred to as 
``precision term'' and ``accuracy term'', respectively.

\begin{figure}[!h]
  \centering
  \includegraphics[width=16cm]{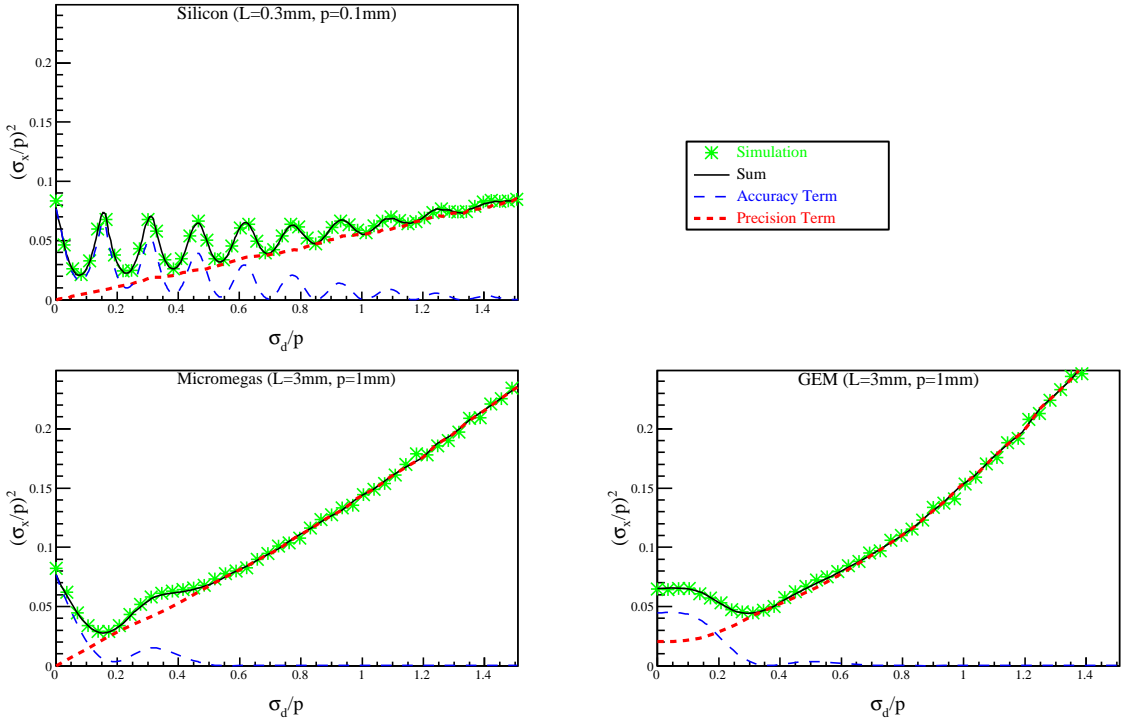}
  \caption{Comparison between our simulation (green, star) and the sum (black, solid line) of the 
  precision term (red, dotted line) and the accuracy term (blue, dashed line) for three different detector models.
  The precision term and the accuracy term were numerically computed here.
  }
  \label{sim_2terms_comparison}
\end{figure}

Each term is separately plotted, together with the simulation results for three detectors in Fig.\,\ref{sim_2terms_comparison}.
The sum of the precision term and the accuracy term matches the corresponding simulation as expected.
Since the precision term increases according to $\sigma_d$, 
the precision term can be recognized as a contribution from the diffusion effect.
On the other hand, the accuracy term has a periodic structure and it is more clearly visible in smaller $\sigma_d$ region, 
especially for the semiconductor detector configuration in which a large number of ionized electrons are produced.
The Micromegas and GEM-based detector configurations give similar results except for small $\sigma_d$ region
because of the additional diffusion in the induction regions.
In Sec.\,\ref{accuracy}, we will revisit the accuracy term which looks more complicated.

\begin{figure}[!h]
  \centering
  \includegraphics[width=10cm]{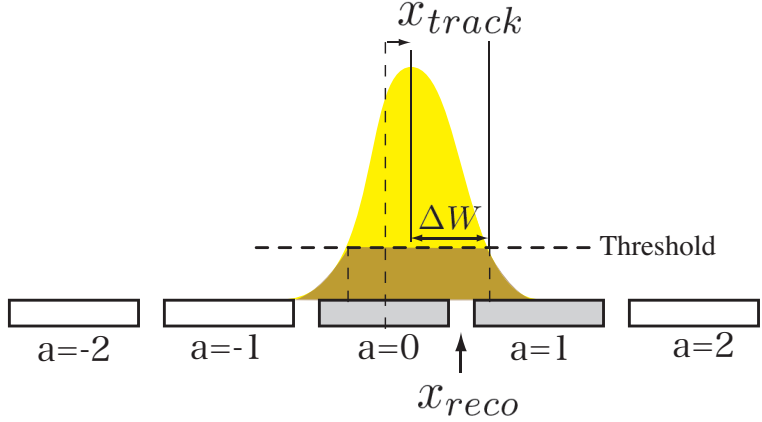}
  \caption{Graphical interpretation of $\Delta W$.
           The yellow shape represents a charge spread.
           $\Delta W$ can be interpreted as an effective width of the charge spread.
           When the strips $a=0$ and $a=1$ collect a certain amount of charges as seen in this figure,
           the reconstructed position ($x_{reco}$) is expected to be between $a=0$ and $a=1$.
           Note that the $\Delta W$ depends not only on the diffusion but also the threshold,
           the track angle, and the diffusion in the amplification region e.g. GEM.
          }
  \label{model}
\end{figure}

\paragraph{Model parameter \boldmath{$\Delta W$} (\boldmath{$\Delta W_{p}:=\Delta W/p$}) and auxiliary parameter $\mu_{\pm n}$}
To investigate more concretely, let us introduce a new parameter representing an effective width of charge spread.
In other words, we assume that the strips within $x_{track}\pm\Delta W$ are fired (Fig.\,\ref{model}).
Note that this $\Delta W$ depends not only on the diffusion but also on the incident track angle as well as the electronic threshold. 
In reality this $\Delta W$ 
varies because of
the stochastic processes such as
the diffusion, the ionization, and the gas gain.
However we will use constant $\Delta W$ for simplicity.

\begin{figure}[!h]
  \centering
  \includegraphics[width=10cm]{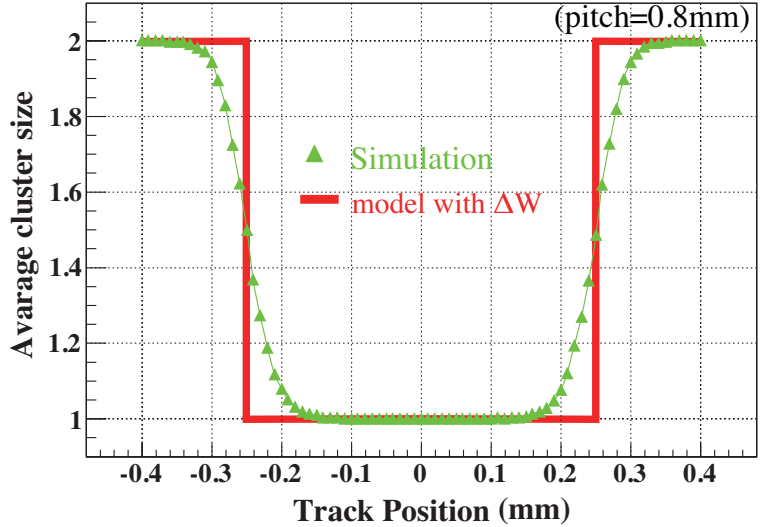}
  \caption{Schematic view to show difference between our simulation and one-parameter model in terms of the cluster size.
  $\Delta W \approx 0.15$~mm in this example.
  }
  \label{sim_model_clustersize}
\end{figure}

\begin{figure}[!h]
  \centering
  \includegraphics[width=14cm]{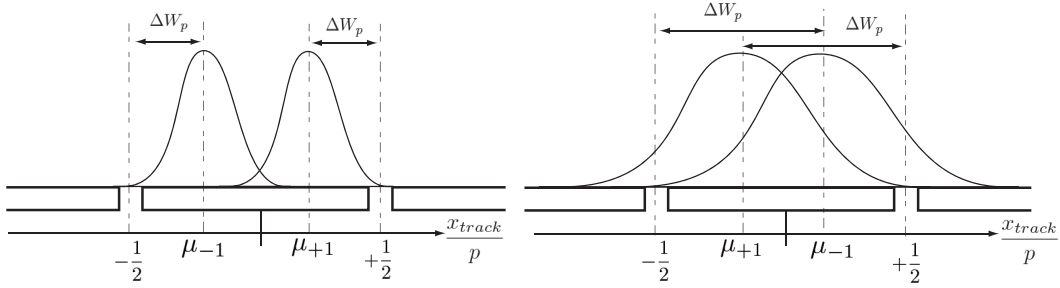}
  \caption{
    The charge spread touches the boundary of the strip $a=1$ ($-1$) at $x_{track}=\mu_{+1}$ ($\mu_{-1}$).
    On the left is a case of $\mu_{+1}>\mu_{-1}$. On the right is a case of $\mu_{-1}>\mu_{+1}$.
  }
  \label{Mu_n}
\end{figure}

As can be seen in Fig.\,\ref{sim_model_clustersize}, 
the cluster size depends on the track position in a readout strip.
To describe the boundary where the cluster size changes, let us define $\mu_{\pm n}$:
\begin{eqnarray}
  \label{Eq:Mu_n}
  \mu_{\pm n} := \pm \left(n - \frac{1}{2} - \frac{\Delta W}{p}\right)
              :=\pm \left(n - \frac{1}{2} - \Delta W_{p}\right),
\end{eqnarray}
where we defined $\Delta W_{p}$ as $\Delta W/p$.
$\mu_{+n}$ is the value of $x_{track}/p$ where the charge spread touches to the strip of $a=n$ from the lower side,
and 
$\mu_{-n}$ is the value of $x_{track}/p$ where the charge spread touches to the strip of $a=-n$ from the upper side.
Fig.\,\ref{Mu_n} shows example cases of $n=1$. 
Note that
\begin{eqnarray}
  \label{xbar_condition}
  \frac{x_{reco}}{p} = \left\{
  \begin{array}{l}
   ~ 0 ~~~ (\mu_{min} \leq x_{track}/p \leq \mu_{max}) \\
   ~ 1 ~~~ (\mu_{max} < x_{track}/p )\\
   -1 ~~ (\mu_{min} > x_{track}/p )
  \end{array}
  \right.
\end{eqnarray}
with $\mu_{max}$ being $\max(\mu_{+1},\mu_{-1})$ and $\mu_{min}$ being $\min(\mu_{+1},\mu_{-1})$.

\subsection{One-parameter model for the accuracy term}
\label{accuracy}
If we consider all the hit strips are in contact with their neighbor hit strips,
the reconstructed position in an event is always discrete:
\begin{eqnarray}
  \frac{x_{reco}}{p} = 0,\pm 0.5, \pm 1.0, \pm 1.5 \cdots
\end{eqnarray}
This fact motivates us to rewrite the accuracy term in the following general expression:
\begin{eqnarray}
  \label{systerm}
  {\rm Accuracy ~ Term} & = & \int_{-1/2}^{+1/2} d\left(\frac{x_{track}}{p}\right)
                            \left(
                                \left<\frac{x_{reco}}{p}\right> - \frac{x_{track}}{p}
                            \right)^2 \cr
                      & \equiv & 
                            \int_{-1/2}^{+1/2} d\left(\frac{x_{track}}{p}\right)
                            \left(
                              \sum_{k=-\infty}^{+\infty}R_k(x_{track}) 
                            \frac{k}{2}-\frac{x_{track}}{p}\right)^2,
\end{eqnarray}
where $R_k(x_{track})$ ($k$:integer) is the probability that the position is reconstructed at $k/2$.
Note that a special condition $R_0=1$, $R_{k\neq0}=0$ gives a well known formula $(\sigma_x/p)^2 = 1/12$.

As a first step,
let us develop Equation~(\ref{systerm})
assuming a specific condition $\Delta W_p\leq 1$ 
in which Equation~(\ref{xbar_condition}) should be satisfied:

\begin{eqnarray}
{\rm Accuracy~Term} &=& 
                            \int_{-1/2}^{+1/2} d\left(\frac{x_{track}}{p}\right)
                            \left(
                              \sum_{k=-\infty}^{+\infty}R_k(x_{track}) 
                            \frac{k}{2}-\frac{x_{track}}{p}\right)^2\cr
                    &=&        \int_{-1/2}^{+1/2} d\left(\frac{x_{track}}{p}\right)
                            \left(
                              R_{+1}(x_{track})\cdot\left(\frac{1}{2}\right)
                             +R_{0}(x_{track})\cdot\left(\frac{0}{2}\right)
                             +R_{-1}(x_{track})\cdot\left(\frac{-1}{2}\right)
                            -\frac{x_{track}}{p}\right)^2\cr
                    &=&        \int_{-1/2}^{+1/2} d\left(\frac{x_{track}}{p}\right)
                            \left(
                              \frac{f_{+1}}{2}-\frac{f_{-1}}{2}
                            -\frac{x_{track}}{p}\right)^2,
\end{eqnarray}
where we defined ``turn-on'' functions of readout strip of $a=\pm 1$ as :
\begin{eqnarray}
f_{+1} := \left\{
         \begin{array}{l}
           1 ~~ (~ \frac{x_{track}}{p}\geq \mu_{+1} ~)\\
           0 ~~ (~ \frac{x_{track}}{p} < \mu_{+1} ~)
         \end{array}\right.\cr
f_{-1} := \left\{
         \begin{array}{l}
           1 ~~ (~ \frac{x_{track}}{p}\leq \mu_{-1} ~)\\
           0 ~~ (~ \frac{x_{track}}{p} > \mu_{-1} ~).
         \end{array}\right.
\end{eqnarray}

\begin{figure}[!ht]
  \centering
  \includegraphics[width=10cm]{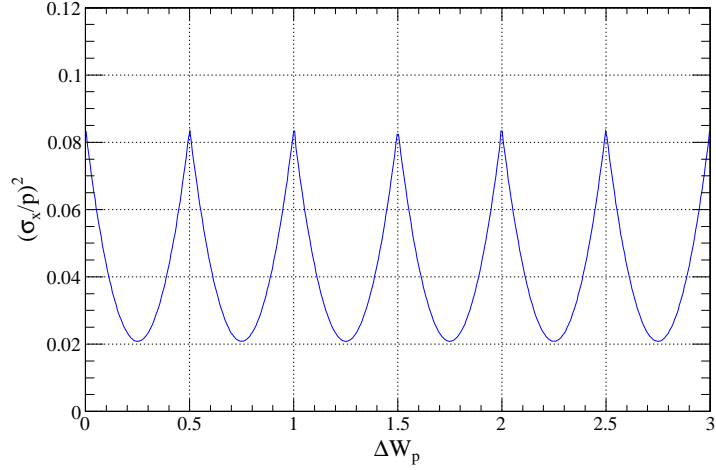}
  \caption{
    The accuracy term described with the one-parameter model (Equation~(\ref{formula_simple})).
    This model explains the wavy structure seen in the accuracy term.
  }
  \label{formula_simple_only}
\end{figure}

This model is only valid within $\Delta W_p \leq 1$ as assumed,
but this can be easily generalized in a similar fashion:
\begin{eqnarray}
\label{formula_simple}
{\rm Accuracy~Term} &=& 
                            \int_{-1/2}^{+1/2} d\left(\frac{x_{track}}{p}\right)
                            \left(
                              \sum_{k=-\infty}^{+\infty}R_k(x_{track}) 
                            \frac{k}{2}-\frac{x_{track}}{p}\right)^2\cr
                    &=&        \int_{-1/2}^{+1/2} d\left(\frac{x_{track}}{p}\right)
                            \left(
                              \sum_{n=1}^{\infty}\left(\frac{f_{+n}}{2}-\frac{f_{-n}}{2}\right)
                            -\frac{x_{track}}{p}\right)^2,
\end{eqnarray}
with
\begin{eqnarray}
\label{fn_simple}
f_{+n} := \left\{
         \begin{array}{l}
           1 ~~ (~ \frac{x_{track}}{p}\geq \mu_{+n} ~)\\
           0 ~~ (~ \frac{x_{track}}{p} < \mu_{+n} ~)
         \end{array}\right.\cr
f_{-n} := \left\{
         \begin{array}{l}
           1 ~~ (~ \frac{x_{track}}{p}\leq \mu_{-n} ~)\\
           0 ~~ (~ \frac{x_{track}}{p} > \mu_{-n} ~).
         \end{array}\right.
\end{eqnarray}
Since the number of hit strips is likely less than 7 in most practical cases,
$f_{+n}$ and $f_{-n}$ for $n>3$ are expected to be $0$.
In such a case the sum in Equation~(\ref{formula_simple}) is necessary only for $n\leq~\sim2$.
A numerical computation of Equation~(\ref{formula_simple})
is shown in Fig.\,\ref{formula_simple_only} to see how the spatial resolution evolves with $\Delta W_p$.
This one-parameter model explains the wavy structure seen in the accuracy term.

\begin{figure}[!ht]
  \centering
  \includegraphics[width=10cm]{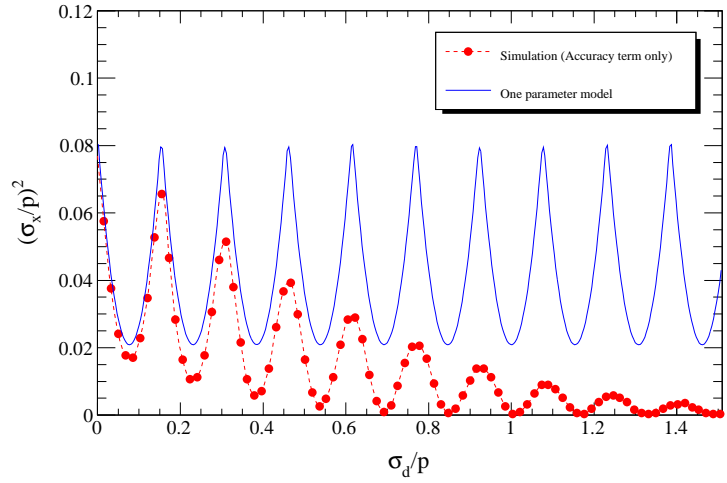}
  \caption{
  The one-parameter model compared to
  the simulation. In this example, we set $\Delta W=3.25\sigma_d$ so that their periodic patterns are matched.}
  \label{sim_model_comparison}
\end{figure}

Fig.\,\ref{sim_model_comparison} shows comparison between
the accuracy term from the simulation and the one from the one-parameter model.
The discrepancy getting larger according to $\sigma_d/p$
is caused by the fact that the constant $\Delta W$ assumption becomes 
no longer valid due to large diffusion with limited ionization statistics.

\subsection{Second model parameter \boldmath{$\delta$} and two-parameter model for the accuracy term}
\label{sec:improvedmodel}
\begin{figure}[!ht]
  \centering
  \includegraphics[width=10cm]{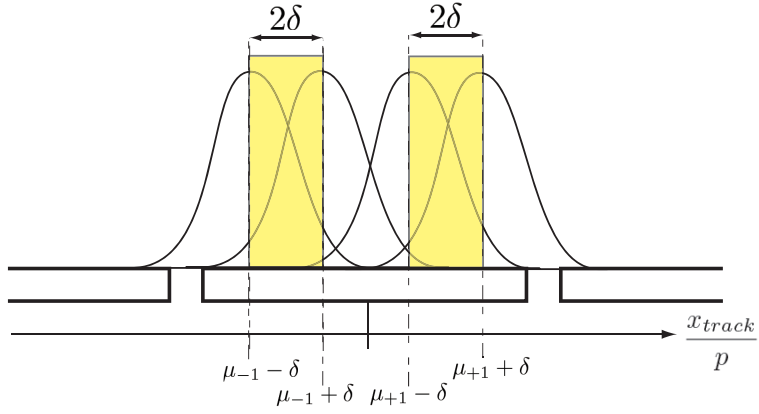}
  \caption{Transition region around $\mu_{\pm 1}$ with its width of $2\delta$ in the two-parameter model.}
  \label{transition}
\end{figure}

\begin{figure}[!ht]
  \centering
  \includegraphics[width=10cm]{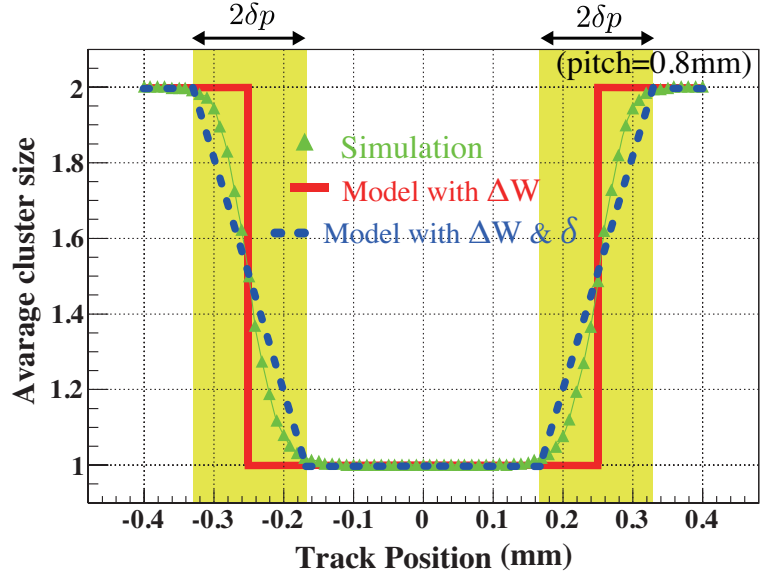}
  \caption{A comparable plot to Fig.\,\ref{sim_model_clustersize}, but with including the transition regions.
The blue dotted line corresponds to the
new model introduced at Sec.\,\ref{sec:improvedmodel}.}
  \label{sim_model_clustersize2}
\end{figure}

The diffusion causes fluctuations in $\Delta W$, however we did not take the fact into account in Sec.\,\ref{accuracy}.
When $\Delta W$ fluctuates,
the boundaries $\mu_n$ defined in Equation~(\ref{Eq:Mu_n}) are no longer constant in each event.
This fact was already observed in Fig.\,\ref{sim_model_clustersize}, where the transition boundary between the cluster size of 1 and that of 2 can be defined as a point (e.g. $\pm$0.25) in the model case (Red line) while it can be only defined as a range (e.g. from $\pm$0.18 to $\pm$0.32) in the realistic simulation (Green line).
In order to take this effect into account,
let us define ``transition regions'' with a width of $2\delta$ as highlighted in yellow in Fig.\,\ref{transition}.
The idea is to linearly weight $f_{\pm n}$ in the transition region.
Considering a finite number of charges,
the transition regions are expected to be enlarged with respect to $\Delta W_{p}$.
We therefore constrain $\delta$ being proportional to $\Delta W_{p}$ as a first approximation. 
Equation~(\ref{fn_simple}) can be modified as follows:
\begin{eqnarray}
\label{fn_delta}
  f_{+n} &=& \left\{
         \begin{array}{l}
           1  ~~~~~~~~~ (~ \frac{x_{track}}{p} > \mu_{+n}+\delta ~) \\
           t_n ~~~~~~~~ (~ \mu_{+n}-\delta \leq \frac{x_{track}}{p} \leq \mu_{+n}+\delta)\\
           0  ~~~~~~~~~ (~ \frac{x_{track}}{p} < \mu_{+n}-\delta ~)
         \end{array},
        \right.\cr
  f_{-n} &=& \left\{
         \begin{array}{l}
           0\quad\;\; ~~~~ (~ \frac{x_{track}}{p} > \mu_{-n}+\delta ~) \\
           s_{n}      ~~ (~\mu_{-n}-\delta \leq \frac{x_{track}}{p} < \mu_{-n}+\delta ~) \\
           1\quad\;\; ~~~~ (~\frac{x_{track}}{p} < \mu_{-n}-\delta ~) \\
         \end{array},
        \right.\cr\cr
  t_n &=& \frac{1}{2\delta}\left( \frac{x_{track}}{p} - \left( \mu_{+n} - \delta \right) \right),\\
  s_n &=& 1-\frac{1}{2\delta}\left( \frac{x_{track}}{p} - \left( \mu_{-n} - \delta \right) \right).
\end{eqnarray}
Note that $\delta$ works as a second parameter to describe the $\Delta W$ fluctuations.
The main difference from the one-parameter model is, that
$f_{\pm n}$ could shift simultaneously depending on the track position $x_{track}$ within the transition region.
This can recover the $\Delta W$ fluctuation effect due to the diffusion in the drift region.
This improvement can also be seen from the viewpoint of the cluster size (Fig.\,\ref{sim_model_clustersize2}).

\begin{figure}[!ht]
  \centering
  \includegraphics[width=10cm]{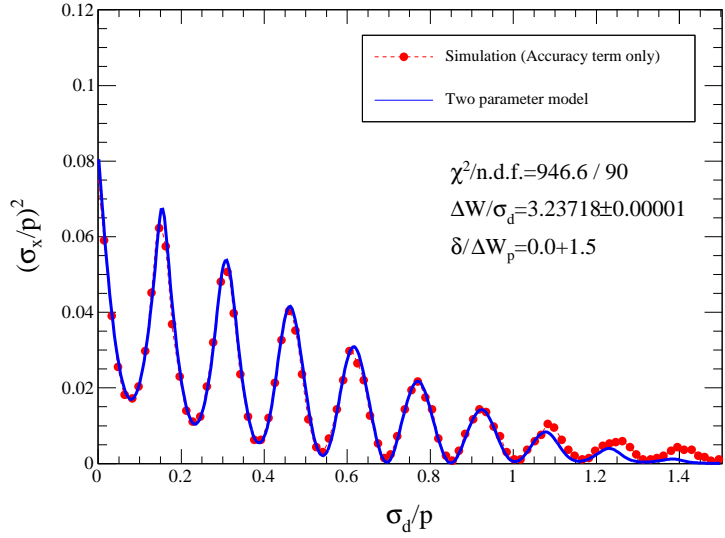}
  \caption{The two-parameter model with fitted parameters compared to the simulation.
  To compute $\chi^2$, a fixed error of $0.0005$ is assumed to each simulation point.
  }
  \label{fitted_with_delta}
\end{figure}

A numerical computation of Equation~(\ref{formula_simple}) with Equation~(\ref{fn_delta}) 
is shown in Fig.\,\ref{fitted_with_delta} to see how the spatial resolution evolves with $\Delta W_{p}$.
This two-parameter model can describe the simulation result well with just two parameters of $\Delta W_{p}$ and $\delta$.

\subsection{Sigmoid expression for two-parameter model}
Although the two-parameter model describes well the simulation result,
it is difficult to generalize the integration over $x_{track}/p$ due to
the condition analysis (e.g. {\it $f_{+n}=1$} when {\it $x_{track}/p\leq\mu_{+n}$}).
To improve this situation
we propose another formulation of $f_{\pm n}$ with the sigmoid functions:
\begin{eqnarray}
\label{fsigmoid}
f_{+n}&=&\frac{1}{1+\exp[-\frac{x_{track}/p-\mu_{+n}}{2\delta}]},\cr
f_{-n}&=&\frac{1}{1+\exp[ \frac{x_{track}/p-\mu_{-n}}{2\delta}]},
\end{eqnarray}
where $\mu_{\pm n}$ are defined in Equation~(\ref{Eq:Mu_n}).
Noting Equation~(\ref{formula_simple}) can be rewritten as:
\begin{eqnarray}
\label{formula_exp}
{\rm Accuracy~Term} 
                    &=&     
                           \int_{-1/2}^{+1/2} d\left(\frac{x_{track}}{p}\right)
                            \left(
                              \frac{1}{4}\sum_{n=1}^{\infty}\left(f_{+n}-f_{-n}\right)^2
                            + \frac{1}{2}\sum_{n<n'}\left(f_{+n}-f_{-n}\right)\left(f_{+n'}-f_{-n'}\right)
                            \right.\cr
                    && ~~~~~~~~~~~~~~~ \left.
                           -\sum_{n=1}^{\infty}\frac{x_{track}}{p}\left(f_{+n}-f_{-n}\right)
                              +\left(\frac{x_{track}}{p}\right)^2
                            \right),
\end{eqnarray}
and using the following equations, one can compute the accuracy term:
\begin{flalign}
   \int_{-1/2}^{+1/2} d\left(\frac{x_{track}}{p}\right)
                       \left(f_{+n}-f_{-n}\right)^2
=&
\left[2\delta\ln\left(\frac{\exp[\frac{x_{track}/p-\mu_{+n}}{2\delta}]+1}{\exp[-\frac{x_{track}/p-\mu_{-n}}{2\delta}]+1}\right) 
- \frac{2\delta}{\exp[-\frac{x_{track}/p-\mu_{+n}}{2\delta}]+1} 
+ \frac{2\delta}{\exp[\frac{x_{track}/p-\mu_{-n}}{2\delta}]+1}
\right.
\cr 
&-\left.
 \frac{4\delta}{\exp[\frac{\mu_{+n}-\mu_{-n}}{2\delta}]-1}
  \ln\left(\frac{\exp[\frac{x_{track}/p-\mu_{-n}}{2\delta}]+1}{\exp[\frac{x_{track}/p-\mu_{+n}}{2\delta}]+1}\right)
\right]_{-1/2}^{+1/2},\cr
\end{flalign}
\begin{flalign}
  &\int_{-1/2}^{+1/2} d\left(\frac{x_{track}}{p}\right)
                       \left(f_{+n}-f_{-n}\right)\left(f_{+n'}-f_{-n'}\right)
=\cr
&~~~~~~1+2\delta
\left[
  \ln \left( \exp\left[-\frac{x_{track}/p-\mu_{+n}}{2\delta}+1\right] \right) 
           \left(\frac{1}{\exp\left[-\frac{\mu_{-n'}-\mu_{+n}}{2\delta}\right]-1} - \frac{1}{\exp\left[\frac{\mu_{+n'}-\mu_{+n}}{2\delta}\right]-1}\right) 
\right.\cr
&\left.
~~~~~~~~~~~~~ + \ln \left( \exp\left[-\frac{x_{track}/p-\mu_{-n}}{2\delta}+1\right] \right) 
           \left(\frac{1}{\exp\left[-\frac{\mu_{-n'}-\mu_{-n}}{2\delta}\right]-1} - \frac{1}{\exp\left[\frac{\mu_{+n'}-\mu_{-n}}{2\delta}\right]-1}\right) 
\right.\cr
&\left.
~~~~~~~~~~~~~ + \ln \left( \exp\left[-\frac{x_{track}/p-\mu_{+n'}}{2\delta}+1\right] \right) 
           \left(\frac{1}{\exp\left[\frac{\mu_{+n'}-\mu_{-n}}{2\delta}\right]-1} - \frac{1}{\exp\left[-\frac{\mu_{+n'}-\mu_{+n}}{2\delta}\right]-1}\right) 
\right.\cr
&\left.
~~~~~~~~~~~~~ + \ln \left( \exp\left[-\frac{x_{track}/p-\mu_{-n'}}{2\delta}+1\right] \right) 
           \left(\frac{1}{\exp\left[\frac{\mu_{-n'}-\mu_{-n}}{2\delta}\right]-1} - \frac{1}{\exp\left[-\frac{\mu_{-n'}-\mu_{+n}}{2\delta}\right]-1}\right) 
\right]_{-1/2}^{+1/2},\cr
\end{flalign}

\begin{flalign}
   \int_{-1/2}^{+1/2} d\left(\frac{x_{track}}{p}\right)
                           \frac{x_{track}}{p}\left(f_{+n}-f_{-n}\right)
=&
2\delta
\left[ 
\frac{x_{track}}{p}\ln\left\{\left(\exp\left[\frac{x_{track}/p-\mu_{+n}}{2\delta}\right]+1\right)
\left(\exp\left[-\frac{x_{track}/p-\mu_{-n}}{2\delta}\right]+1\right)\right\}\right.\cr
+& 
\left. 
2\delta \left\{ 
{\rm DiLog}\left(-\exp\left[\frac{x_{track}/p-\mu_{+n}}{2\delta}\right]\right)
-{\rm DiLog}\left(-\exp\left[-\frac{x_{track}/p-\mu_{-n}}{2\delta}\right]\right)
\right\}
\right]_{-1/2}^{+1/2},\cr
\end{flalign}
where DiLog is defined as
-$\int_0^{x}\frac{\ln(1-t)}{t}dt$,
and implemented in {\it ROOT} framework\cite{Brun:1997pa,dilog},
and
\begin{flalign}
   \int_{-1/2}^{+1/2} d\left(\frac{x_{track}}{p}\right)
        \left(\frac{x_{track}}{p}\right)^2
= \frac{1}{12}.
\end{flalign}

\begin{figure}[!ht]
  \centering
  \includegraphics[width=10cm]{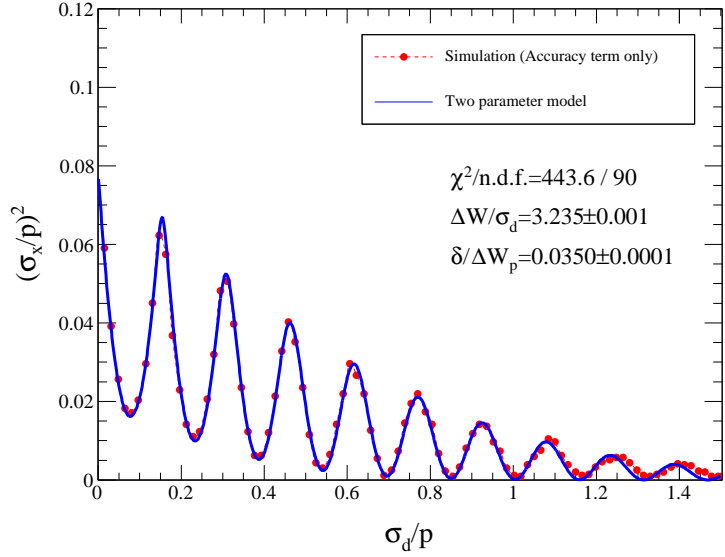} 
  \caption{
  The two-parameter model with sigmoid functions compared to the simulation for silicon detector configuration.
  To compute $\chi^2$, a fixed error of $0.0005$ is assumed to each simulation point.
  }
  \label{fit_with_exp_si}
\end{figure}

\begin{figure}[!ht]
  \centering
  \includegraphics[width=10cm]{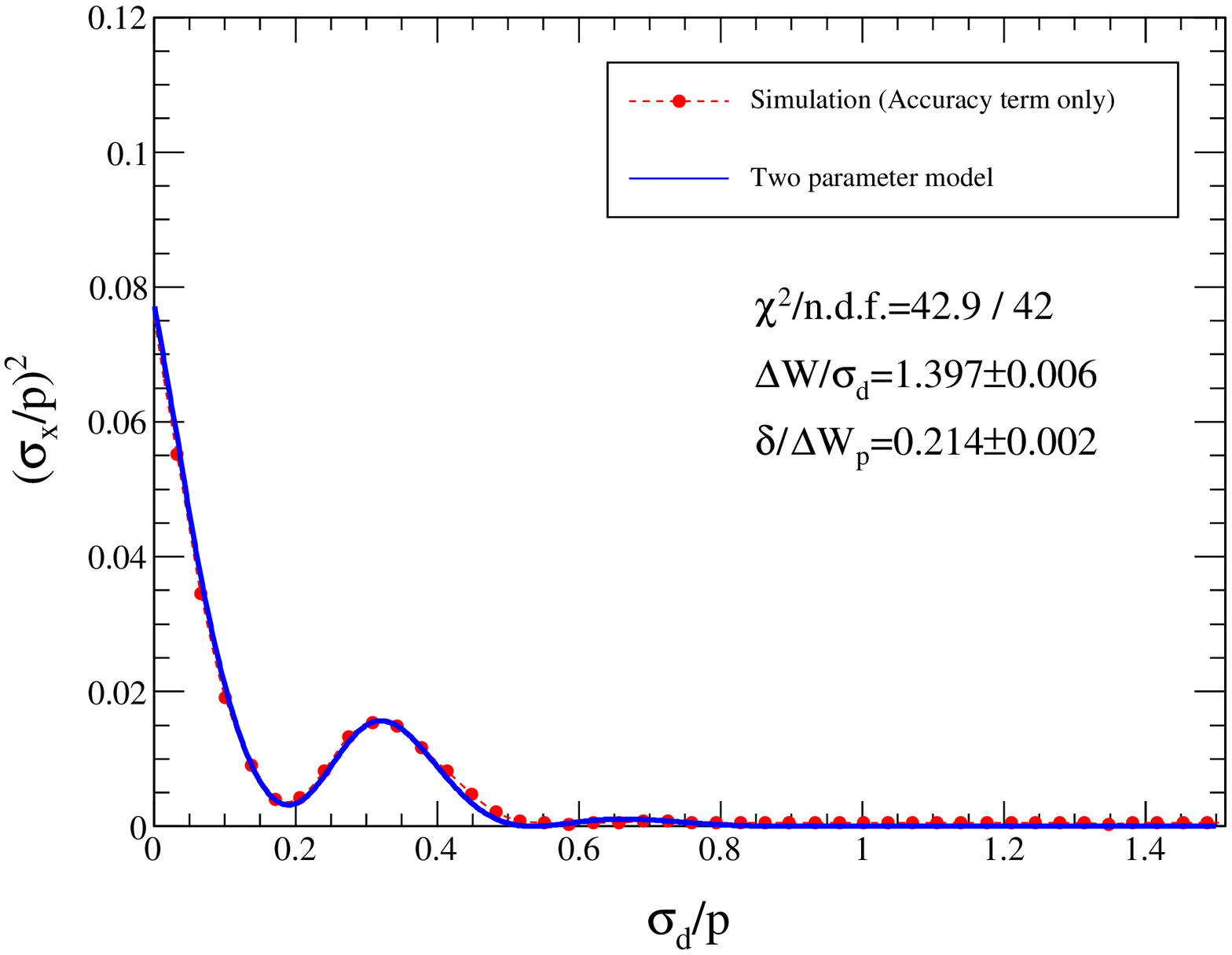} 
  \caption{
  The two-parameter model with sigmoid functions compared to the simulation for Micromegas-based detector configuration.
  To compute $\chi^2$, a fixed error of $0.0005$ is assumed to each simulation point.
  }
  \label{fit_with_exp_mm}
\end{figure}

\begin{figure}[!ht]
  \centering
  \includegraphics[width=10cm]{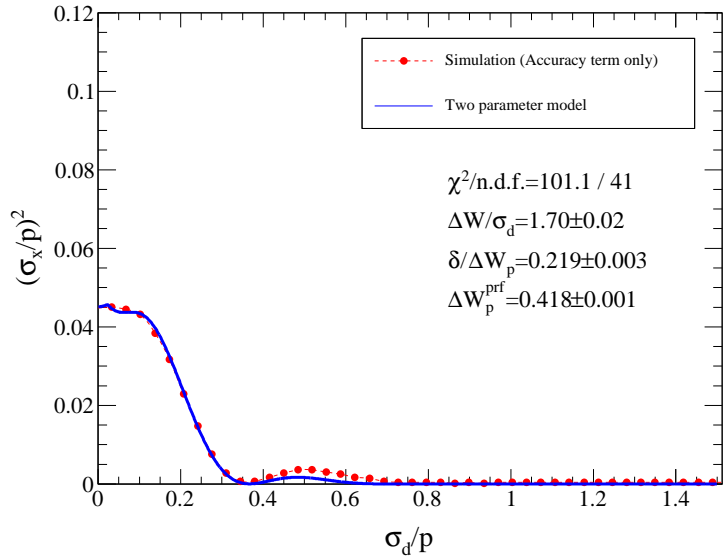} 
  \caption{
  The two-parameter model with sigmoid functions compared to the simulation for GEM-based detector configuration.
  To compute $\chi^2$, a fixed error of $0.0005$ is assumed to each simulation point.
  }
  \label{fit_with_exp_gem}
\end{figure}

Figs.\,\ref{fit_with_exp_si}, \ref{fit_with_exp_mm}, and \ref{fit_with_exp_gem} show
fitting results with using the sigmoid expression model for a silicon detector case,
a Micromegas-based detector case and a GEM-based detector case. 
For GEM-based detectors, an additional parameter $\Delta W^{prf}_{p}$ must be introduced to consider
the diffusion in the induction region,
and $\mu_{\pm n}$ in Equation~(\ref{Eq:Mu_n}) is modified as:
\begin{eqnarray}
  \label{Eq:Mu_n_GEM}
  \mu_{\pm n} := \pm \left(n - \frac{1}{2} - \sqrt{\left(\Delta W_p\right)^2+\left(\Delta W^{prf}_p\right)^2}\right).
\end{eqnarray}


\subsection{Cluster size}
\begin{figure}[!ht]
  \centering
  \includegraphics[width=10cm]{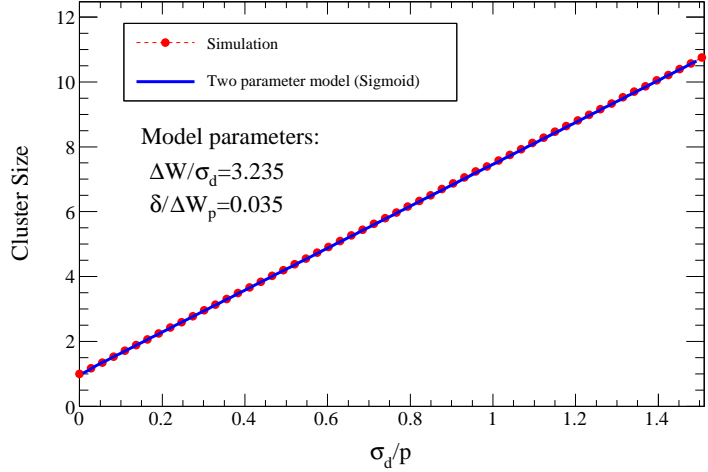} 
  \caption{
  Cluster size compared with the two-parameter model with sigmoid functions and the simulation for silicon detector configuration.
  The parameters for the model are same as the ones obtained by fitting the spatial resolution.
  }
  \label{Fig:formula_clsize}
\end{figure}

The cluster size is defined as the average number of hit strips.
Using $f_{\pm n}$ defined in the previous sections, 
the cluster size is expressed by
\begin{flalign}
\label{eq:clsize}
  {\rm Cluster ~~ Size} = 1 + \int_{-1/2}^{+1/2}d\left(\frac{x_{track}}{p}\right)
                              \left\{
                                       \sum_{n=1}^{\infty} \left( f_{+n} + f_{-n} \right)
                              \right\}.
\end{flalign}
If adopting the sigmoid definition of Equation~(\ref{fsigmoid}), it can be rewritten as follows:
\begin{flalign}
\label{Eq:formula_clustersize}
{\rm Cluster ~~ Size} = 1+ \sum_{n=1}^{\infty}
              \left\{
                  2+ 2\delta\left[ 
                               \ln\left(\exp\left[-\frac{x_{track}/p-\mu_{+n}}{2\delta}\right]+1\right)
                              -\ln\left(\exp\left[\frac{x_{track}/p-\mu_{-n}}{2\delta}\right]+1\right)
                           \right]^{+1/2}_{-1/2}
             \right\}.\cr
\end{flalign}
An example plot of Equation(\ref{Eq:formula_clustersize}) is shown in Fig.\,\ref{Fig:formula_clsize} together with the simulation result.
The result shows that the two-parameters can describe the cluster size as well as the spatial resolution.

\subsection{An idea for optimizing the spatial resolution}
If the spatial resolution is dominated by the accuracy term, 
and once the spatial resolution and the cluster size are measured, 
one can obtain the two model parameters ($\Delta W_{p}$ and $\delta$)
from an experiment
by solving 
the simultaneous equations of (\ref{formula_exp}) and (\ref{Eq:formula_clustersize}).
As seen in Figs.\,\ref{fit_with_exp_si}, \ref{fit_with_exp_mm}, and \ref{fit_with_exp_gem}, 
the two-parameter model will give the potential performances of the spatial resolution for detectors,
and thus will give a guiding principle to optimize the spatial resolution for instance
by changing electronics threshold, the electric field, gas mixture or readout pitch.

To demonstrate this new idea of the optimization, 
suppose a silicon detector characterized by $(\sigma_{x}/p)^2=0.0655$
and the cluster size of 1.975.
Here we took specific numbers from Figs.\,\ref{fit_with_exp_si} and \ref{Fig:formula_clsize},
which correspond to the values at $\sigma_{d}/p=0.15$ respectively,
so that we can validate findings by comparing with the simulation results.
Practically these two values are expected to be measured in an experiment.
$\Delta W_p = 0.487$ and $\delta=0.0164$ were obtained by numerically solving
the simultaneous equations of (\ref{formula_exp}) and (\ref{Eq:formula_clustersize}).

\begin{figure}[h]
  \centering
  \includegraphics[width=10cm]{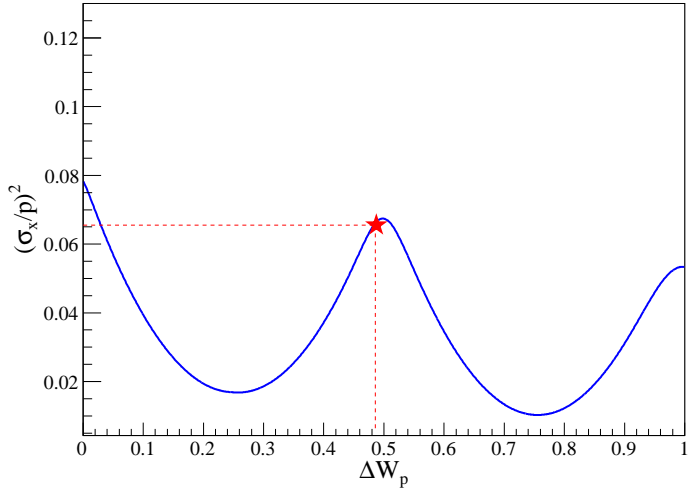} 
  \caption{A model given by $\delta/\Delta W = 0.0337$.
  The blue line shows potential performances and the red star
  indicates the original performance $(\sigma_{x}/p)^2=0.0655$
  which was used as one of the inputs.}
  \label{model_reco}
\end{figure}

A model given by $\delta/\Delta W_p = 0.0337$ predicts potential evolutions of the spatial resolution with respect to
$\Delta W_{p}$, which is drawn in the blue line in Fig.\,\ref{model_reco}.
The original point $\Delta W_{p}=0.487$ is highlighted in the red star in the same figure.
In this example, the condition is not optimized at all in terms of the spatial resolution,
which means the model indicates possibilities to improve the spatial resolution by changing $\Delta W_{p}$.
This result is consistent with the simulation in Fig.\,\ref{fit_with_exp_si}.

\section{Conclusion}
\label{sec:conclusion}
From our simulation, we found that the spatial resolution with the binary readout has characteristic
behaviors.
In some cases the spatial resolution with the binary readout is better than the one with the analogue readout.
To obtain a better understanding on the spatial resolution with the binary readout, 
we first defined the accuracy term and the precision term that can be separately computed.
Secondly we introduced the one-parameter model that uses $\Delta W_p$ to explain
the wavy structure of the accuracy term and 
the two-parameter model that uses an additional parameter named $\delta$ as well as
$\Delta W_p$ to take into account fluctuations of $\Delta W_p$.
The two-parameter model is applicable for silicon detectors and Micromegas-based detectors.
With one more additional parameter, the model can be also used for GEM-based detectors.
Once the two parameters for a silicon detector or a Micromegas-based detector are found by measuring the spatial resolution and the cluster size,
one can obtain an overview of the potential performances by building a two-parameter model for the detector.
Therefore the models can give an idea to improve the spatial resolution by optimizing
the detector gap size, the strip pitch, electric field in the drift region, comparator threshold, etc. 
For GEM-based detectors, an additional measurement is required to determine the diffusion effect in the induction region.

\section*{Acknowledgement}
We gratefully acknowledge the support of F.R.S.-FNRS (Belgium).

\section*{References}
\bibliographystyle{unsrt}
\bibliography{bibliography}
	 
\end{document}